\documentclass[12pt]{iopart}
\usepackage{epsfig}
\usepackage{bm}

\newcommand{\be}{\begin{eqnarray}}
\newcommand{\ee}{\end{eqnarray}}
\newcommand{\ba}{\begin{array}}
\newcommand{\ea}{\end{array}}

\begin{document}

\title{Quantum lower and upper speed limits}
\author{Kazutaka Takahashi}
\address{Institute of Innovative Research, Tokyo Institute of Technology, 
Kanagawa 226--8503, Japan}

\vspace{10pt}
\begin{indented}
\item[]\today
\end{indented}

\begin{abstract}
We derive generalized quantum speed limit inequalities that 
represent limitations on the time evolution of quantum states.
They are extensions of the original inequality and are applied to 
the overlap between the time-evolved state and an arbitrary state.
We can discuss the lower limit of the overlap, in addition to 
the upper limit as in the original inequality, which allows us 
to estimate the minimum time for the evolution toward a target state.
The inequalities are written by using an arbitrary reference state 
and are flexibly used to obtain a tight bound.
We demonstrate these properties by using 
the twisted Landau--Zener model, the Grover Hamiltonian, and 
a periodically-oscillating Hamiltonian.
\end{abstract}

\section{Introduction}

The quantum speed limit is an inequality that is applied to 
a distance measure of quantum states.
In the Mandelstam--Tamm relation~\cite{MT, F, B, V}
the distance measure is bounded from above by the energy dispersion.
In the Margolus--Levitin relation~\cite{ML}, 
it is bounded by the average energy.
The inequalities are interpreted as time-energy uncertainty relations
and are closely related to the geometric structure of Hilbert space.

The original inequalities applied to quantum pure states can be 
generalized to other problems  
such as quantum mixed states with dissipative environments 
and probability distributions 
in classical stochastic processes~\cite{dCEPH, PCCAS, SCMdC, OO, SFS, FSS}.
Different forms of the speed limit are obtained in different situations, 
but they are written by the corresponding metric and 
can be treated in a unified way.
They are applied to various problems including
quantum control, quantum computing, information processing, 
and stochastic thermodynamics.
We can find many applications in literature~\cite{DC}.

Although the inequality is applied to a broad range of systems, 
it gives a loose bound in most of realistic systems.
The bound usually goes to infinity when we consider  
a large processing time or a large system size.

A different type of quantum speed limit was discussed in recent studies.
The standard limit is applied to the overlap between 
the time-evolved state and the initial state.
An inequality was derived for the overlap 
between the time-evolved state and the adiabatic state in \cite{ST} 
and between different time-evolved states in \cite{HT}.
Some applications of the method can be found in \cite{FLN, Hatomura}.
These results imply that the time evolution of a quantum state is 
characterized in multiple ways by using various reference states.
In this paper, we pursue this problem and derive quantum speed limit 
inequalities satisfied among three quantum states.
The result is a generalization of the original Mandelstam--Tamm relation and 
can be applicable to any quantum states.

This paper is organized as follows.
In section \ref{triqsl}, we derive our main result, 
the quantum upper and lower speed limits.
We study several examples in the following sections.
We treat the twisted Landau--Zener model in section \ref{tlz}, 
the Grover Hamiltonian in section \ref{grover}, and 
a periodically-oscillating Hamiltonian in section~\ref{floquet}.
The result is summarized in section \ref{summary}.

\section{Speed limits from triangle inequality}
\label{triqsl}

We consider quantum states described by the density operator $\hat{\rho}$.
Different quantum states are distinguished by introducing 
a distance measure $D(\hat{\rho}_1,\hat{\rho}_2)$. 
Its properties are prescribed by the standard requirements:
nonnegativity, symmetry, and triangle inequality
\be
 D(\hat{\rho}_1,\hat{\rho}_2)\le 
 D(\hat{\rho}_2,\hat{\rho}_3)+D(\hat{\rho}_3,\hat{\rho}_1). \label{tri}
\ee
It is well-known that 
the trace distance and the fidelity satisfy the requirements and are used 
in many applications~\cite{NC}.
In the present paper, we mainly use the fidelity 
\be
 \Theta(\hat{\rho}_1,\hat{\rho}_2)
 =\arccos{\rm Tr}\,
 \left(\hat{\rho}_1^{1/2}\hat{\rho}_2\hat{\rho}_1^{1/2}\right)^{1/2}.
\ee

When the system is described by the pure state 
$\hat{\rho}=|\psi\rangle\langle\psi|$, the fidelity is written 
by the state overlap
\be
 \Theta_{\psi_1,\psi_2}=\arccos |\langle\psi_1|\psi_2\rangle|.
\ee
Then, the triangle inequality reads
\be
 |\Theta_{\psi_1\psi_3}-\Theta_{\psi_2,\psi_3}| \le \Theta_{\psi_1,\psi_2} 
 \le \Theta_{\psi_1,\psi_3}+\Theta_{\psi_2,\psi_3}.
\ee
We can use this inequality to find a lower limit and a upper limit 
for $\Theta_{\psi_1,\psi_2}$.
The triangle inequality among three quantum states was used to 
derive a novel type of speed limit~\cite{LGC, CC}.
Here, we derive different limits by using the method 
in \cite{ST, HT}.

In most of the applications, we are interested in the fidelity 
between the time-evolved state $|\psi(t)\rangle$ and a target state 
$|\psi_{\rm target}\rangle$.
The former state is obtained from the Schr\"odinger equation 
$i\partial_t|\psi(t)\rangle=\hat{H}(t)|\psi(t)\rangle$
with the Hamiltonian $\hat{H}(t)$.
We also introduce a reference state $|\psi_{\rm ref}(t)\rangle$ 
to write the relation
\be
 \Theta_{\psi_{\rm target},\psi_{\rm ref}(t)}
 -\Theta_{\psi_{\rm ref}(t),\psi(t)}
 \le \Theta_{\psi_{\rm target},\psi(t)}
 \le \Theta_{\psi_{\rm target},\psi_{\rm ref}(t)}+\Theta_{\psi_{\rm ref}(t),\psi(t)}. 
\ee
In the standard quantum speed limit inequality, 
the fidelity $\Theta_{\psi(0),\psi(t)}$ is bounded from above
as $\Theta_{\psi(0),\psi(t)}\le\int_0^t\rmd t'\sigma[\hat{H}(t'),\psi(t')]$ where 
$\sigma[\hat{H},\psi]=(\langle\psi|\hat{H}^2|\psi\rangle
 -\langle\psi|\hat{H}|\psi\rangle^2)^{1/2}$.
The main idea of the present paper is 
to use the inequality for the fidelity $\Theta_{\psi_{\rm ref}(t),\psi(t)}$
developed in \cite{ST, HT}.
To apply the standard Mandelstam--Tamm relation, 
we introduce the Hermitian operator $\hat{H}_{\rm ref}(t)$ from the relation 
$i\partial_t|\psi_{\rm ref}(t)\rangle=\hat{H}_{\rm ref}(t)|\psi_{\rm ref}(t)\rangle$
with $|\psi_{\rm ref}(0)\rangle=|\psi(0)\rangle$.
We write the overlap $\langle\psi_{\rm ref}(t)|\psi(t)\rangle$ 
as $\langle\psi(0)|\tilde{\psi}(t)\rangle$ 
where $|\tilde{\psi}(t)\rangle$ satisfies the Schr\"odinger equation with 
an effective Hamiltonian comprised of 
$\hat{H}(t)$ and $\hat{H}_{\rm ref}(t)$~\cite{ST, HT}.
Then, we obtain 
\be
 \Theta_{\rm l}\le\Theta_{\psi_{\rm target},\psi(t)}\le\Theta_{\rm u}, \label{qsl}
\ee
 where
\be
 && \Theta_{\rm l}=\Theta_{\psi_{\rm target},\psi_{\rm ref}(t)}
 -\int_0^t \rmd t'\,\sigma[\hat{H}(t')-\hat{H}_{\rm ref}(t'),\psi_{\rm ref}(t')], 
 \label{thetal}\\
 && \Theta_{\rm u}=
 \Theta_{\psi_{\rm target},\psi_{\rm ref}(t)}
 +\int_0^t \rmd t'\,\sigma[\hat{H}(t')-\hat{H}_{\rm ref}(t'),\psi_{\rm ref}(t')].  
 \label{thetau}
\ee
This is the main result of the present paper.

The main advantage of the relation in equation (\ref{qsl}) is that 
the inequalities hold for arbitrary choices of a target state 
$|\psi_{\rm target}\rangle$ and a reference state $|\psi_{\rm ref}(t)\rangle$.
By using the reference state $|\psi_{\rm ref}(t)\rangle$, we can derive 
a lower limit of $\Theta_{\psi_{\rm target},\psi(t)}$
in addition to the upper limit.
The bounds can be estimated 
without knowing the time-evolved state $|\psi(t)\rangle$.

We note that the inequalities also hold when we use $|\psi(t')\rangle$ 
in place of $|\psi_{\rm ref}(t')\rangle$ for the variance $\sigma$~\cite{ST}.
The standard upper speed limit is obtained by setting 
$|\psi_{\rm target}\rangle=|\psi(0)\rangle$, $\hat{H}_{\rm ref}(t)=0$, and 
$\sigma[\hat{H}(t')-\hat{H}_{\rm ref}(t'),\psi_{\rm ref}(t')]
\to\sigma[\hat{H}(t'),\psi(t')]$.

Since the fidelity distance satisfies $0\le\Theta\le \pi/2$, 
the bound is useful only when it is within the domain of definition.
When we choose $|\psi_{\rm target}\rangle=|\psi(0)\rangle$, 
$\Theta_{\psi(0),\psi(t)}$ is upper-bounded for a short period of time 
and we can discuss how fast the time-evolved state deviates 
from the initial state.
This gives a result similar to the standard speed limit.
By choosing the reference state in a proper way, we can obtain a tight bound.
On the other hand, when the state evolves toward 
a target state $|\psi_{\rm target}\rangle$ 
with $\langle\psi_{\rm target}|\psi(0)\rangle=0$,
we can estimate a minimum time $t$ 
for the time-evolved state $|\psi(t)\rangle$
to reach the target state as $|\langle\psi_{\rm target}|\psi(t)\rangle|=1$.

There are several ways to utilize the bounds.
In the following sections, 
we discuss possible applications by using several model Hamiltonians.

\section{Twisted Landau--Zener model: tight bound at large time}
\label{tlz}

When we exactly know the state evolution $|\psi_{\rm ref}(t)\rangle$
under the Hamiltonian $\hat{H}_{\rm ref}(t)$, we can estimate the bound 
on the unknown state $|\psi(t)\rangle$ 
under the Hamiltonian $\hat{H}(t)$.
Since the bound is represented by the time integration of 
the variance $\sigma$,
the bound becomes tight when 
$\sigma$ takes nonzero values only for a finite interval of $t$.

We study these properties by using 
the twisted Landau--Zener Hamiltonian~\cite{Berry90} 
\be
 \hat{H}(t)= \left(\ba{cc} vt & \Delta \rme^{-i\varphi(t)} \\ 
 \Delta\rme^{i\varphi(t)} & -vt \ea\right).
\ee
$\Delta$ and $v$ are arbitrary positive parameters
and $\varphi(t)$ represents an arbitrary function of $t$.
This Hamiltonian is compared to the standard form of the Landau--Zener
Hamiltonian~\cite{Landau, Zener}
\be
 \hat{H}_{\rm ref}(t)= \left(\ba{cc} vt & \Delta \\ 
 \Delta & -vt \ea\right). \label{LZ}
\ee
We start the time evolution from 
\be
 |\psi(-\infty)\rangle=\left(\ba{c} 1 \\ 0 \ea\right),
\ee
and study the overlap $\langle\psi_{\rm target}|\psi(\infty)\rangle$ where 
$|\psi_{\rm target}\rangle=|\psi(-\infty)\rangle$.
$|\psi_{\rm ref}(t)\rangle$ is exactly solvable and 
we obtain the Landau--Zener formula 
\be
 |\langle\psi_{\rm target}|\psi_{\rm ref}(\infty)\rangle|
 =\exp\left(-\frac{\pi\Delta^2}{2v}\right).
\ee

The twisted Landau--Zener model is equivalent to the 
Landau--Zener model with a nonlinear protocol.
It is shown by using the transformation to the rotating frame as 
\be
 &&\exp\left[\frac{i}{2}\varphi(t)\left(\ba{cc} 1 & 0 \\ 
 0 & -1 \ea\right)\right]\left(\hat{H}(t)-i\partial_t\right)
 \exp\left[-\frac{i}{2}\varphi(t)
 \left(\ba{cc} 1 & 0 \\ 0 & -1 \ea\right)\right] \nonumber\\
 &=& \left(\ba{cc} vt-\frac{1}{2}\dot{\varphi}(t) & \Delta \\ 
 \Delta & -\left(vt-\frac{1}{2}\dot{\varphi}(t)\right) \ea\right),
\ee
where the dot symbol denotes the time derivative.
It is difficult to obtain the general solution for a given $\varphi(t)$ 
and we apply the inequalities derived in the previous section.

With the use of the reference Hamiltonian in equation (\ref{LZ}), 
the bounds of $\Theta_{\psi_{\rm target},\psi(\infty)}$ 
in the twisted Landau--Zener model are written as 
\be
 \Theta_{\rm u,l}
 = \arccos\exp\left(-\frac{\pi\Delta^2}{2v}\right)
 \pm\int_0^\infty \rmd t\,\sigma(\hat{H}(t)-\hat{H}_{\rm ref}(t),\psi_{\rm ref}(t)).
\ee
Using the relation 
$\sigma(\hat{X},\psi_{\rm ref})\le 
\sqrt{\langle\psi_{\rm ref}|\hat{X}^2|\psi_{\rm ref}\rangle}$,
we can estimate the second term without knowing $|\psi_{\rm ref}(t)\rangle$ as 
\be
 \int_0^\infty \rmd t\,\sigma(\hat{H}(t)-\hat{H}_{\rm ref}(t),\psi_{\rm ref}(t))
 \le 2\Delta \int_0^\infty \rmd t\,\left|\sin\frac{\varphi(t)}{2}\right|.
\ee
To obtain a finite value of the bounds, 
we require that $\varphi(t)$ (mod $2\pi$)  
is nonzero for a finite domain of $t$.
We choose two types of the protocol:
\be
 \varphi(t)=\left\{\ba{ll}
 \pi\left(1+\tanh\frac{t}{\tau}\right) & \mbox{protocol 1}\\
 \pi\exp\left(-\frac{t^2}{\tau^2}\right) & \mbox{protocol 2}
\ea\right..
\ee

\begin{figure}[t]
\begin{center}
\includegraphics[width=0.60\columnwidth]{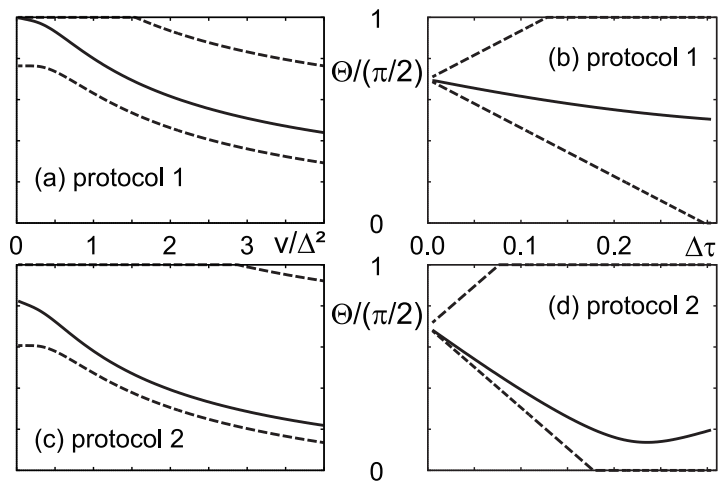}
\caption{
Quantum speed limit in the twisted Landau--Zener model. 
$\Theta_{\psi_{\rm target},\psi(\infty)}$ (solid line) 
and $\Theta_{\rm u,l}$ (dashed lines) 
are plotted as a function of 
parameters $\Delta\tau$ and $v/\Delta^2$.
Each panel shows the result 
with the protocol 1 and $\Delta\tau=0.1$ (a), 
with the protocol 1 and $v/\Delta^2=2.0$ (b), 
with the protocol 2 and $\Delta\tau=0.1$ (c), and
with the protocol 2 and $v/\Delta^2=2.0$ (d).
}
\label{fig01}
\end{center}
\end{figure}

We plot the parameter dependence of $\Theta_{\psi_{\rm target},\psi(\infty)}$
in figure \ref{fig01}.
We can find finite bounds even at the limit $t\to\infty$.
We note that the standard quantum speed limit does not give 
any meaningful result since the bound goes to infinity.

\begin{figure}[t]
\begin{center}
\includegraphics[width=0.50\columnwidth]{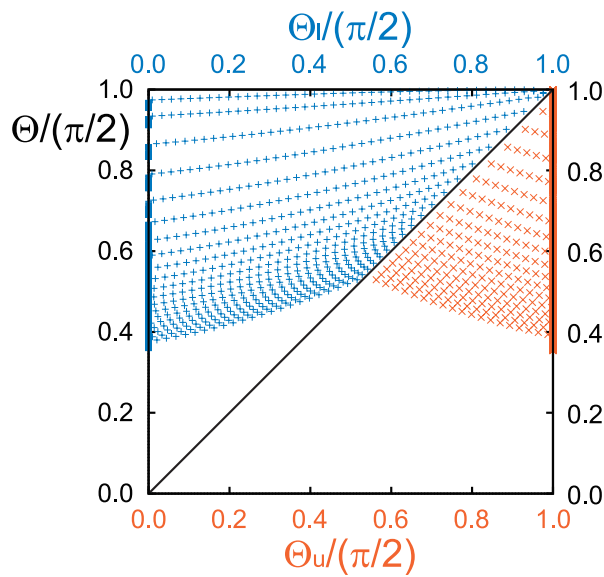}
\caption{
$(\Theta_{\rm u},\Theta_{\psi_{\rm target},\psi(\infty)})$ (blue cross) and 
$(\Theta_{\rm u},\Theta_{\psi_{\rm target},\psi(\infty)})$ (red slanted cross)
in the twisted Landau--Zener model.
The parameters are sampled 
within $0<\Delta\tau\le 1$ and $0<\frac{v}{\Delta^2}\le 4$.
}
\label{fig02}
\end{center}
\end{figure}

We also plot results for various choices of parameters 
in figure \ref{fig02}.
The result shows that the bounds become tight for small $\tau$ 
and large $v$.
The former condition, small $\tau$, is reasonable since 
$\sigma$ becomes small in that limit.
The latter condition represents nonadiabatic regime.
The opposite limit, small $v$, is basically described 
by the adiabatic approximation. 
Our inequalities can be useful in the nonadiabatic regime
rather than in the adiabatic one.


\section{Grover Hamiltonian: protocol-independent bound}
\label{grover}

As we can see from the example in the previous section, 
the bound is in principle dependent on the Hamiltonian and on the protocol.
In the present section, we discuss that a universal bound is obtained by 
choosing the reference state in a proper way.

One of the reasonable choice of the reference state is 
the adiabatic state $|\psi_{\rm ad}(t)\rangle$.
It is written by using the instantaneous eigenstates of 
the Hamiltonian $\hat{H}(t)$.
It corresponds to choosing the reference Hamiltonian 
as the counterdiabatic Hamiltonian
$\hat{H}_{\rm ref}(t)=\hat{H}(t)+\hat{H}_{\rm cd}(t)$.

The counterdiabatic driving is used in the method of shortcuts to 
adiabaticity~\cite{DR03, DR05, Berry09, CRSCGM, STA13, STA19}.
When the Hamiltonian is written by the spectral representation as
\be
\hat{H}(t)=\sum_{n}\epsilon_n(\bm{\lambda}(t))
 |n(\bm{\lambda}(t))\rangle\langle n(\bm{\lambda}(t))|,
\ee 
the counterdiabatic term is given by 
\be
 \hat{H}_{\rm cd}(t)=\dot{\bm{\lambda}}(t) \cdot i\sum_n 
 \left(1-|n(\bm{\lambda}(t))\rangle \langle n(\bm{\lambda}(t))|\right)
 |\bm{\nabla}_{\bm{\lambda}} n(\bm{\lambda}(t))\rangle\langle n(\bm{\lambda}(t))|,
\ee
where $\bm{\lambda}(t)=(\lambda_1(t),\lambda_2(t),\dots)$ is a set of
time-dependent parameters.
The solution of the Schr\"odinger equation with the Hamiltonian 
$\hat{H}(t)+\hat{H}_{\rm cd}(t)$ is given by the adiabatic state 
\be
 |\psi_{\rm ad}(t)\rangle
 &=& \sum_n |n(\bm{\lambda}(t))\rangle 
 \langle n(\bm{\lambda}(0))|\psi(0)\rangle \nonumber\\
 && \times\exp\left[
 -i\int_0^t \rmd t'\epsilon_n(\bm{\lambda}(t'))
 -\int_{\bm{\lambda}(0)}^{\bm{\lambda}(t)} \rmd\bm{\lambda}\cdot
 \langle n(\bm{\lambda})|\bm{\nabla}_{\bm\lambda} n(\bm{\lambda})\rangle
\right].
\ee
The counterdiabatic term is written in a form 
$\hat{H}_{\rm cd}(t)=\dot{\bm{\lambda}}(t)\cdot\bm{A}(\bm{\lambda}(t))$
and the adiabatic gauge potential $\bm{A}(\bm{\lambda})$ characterizes  
the geometric property of the system.

We set $|\psi_{\rm ref}(t)\rangle =|\psi_{\rm ad}(t)\rangle$ 
to find a universal bound.
When we choose the initial state as one of eigenstates of $\hat{H}(0)$,
and use a single parameter $\lambda(t)$, 
the time integration of the variance is written as 
\be
 \int_0^t \rmd t' \sigma(\hat{H}_{\rm cd}(t'),\psi_{\rm ad}(t'))
 = \int_{\lambda(0)}^{\lambda(t)} \rmd\lambda\,\sigma_A(\lambda),
\ee
where $\sigma_A(\lambda)$ represents the variance of the adiabatic gauge 
potential $A$ with respect to the eigenstate.
This representation means that 
the bounds are dependent only on 
the initial and final values of the protocol.
When $\bm{\lambda}(t)$ has multiple components, 
the time integration of the variance depends 
only on the path in parameter space $\bm{\lambda}$ 
and is independent of the velocities on the path.

To demonstrate the described general properties, 
we study the Grover Hamiltonian \cite{Grover, FG, RC}
\be
 \hat{H}(t)=
 A(t)\left(1-|+\rangle\langle+|\right)
 +B(t)\left(1-|0\rangle\langle0|\right).
\ee
We want to select the target state $|\psi_{\rm target}\rangle=|0\rangle$ 
from among a set of states $\{|i\rangle\}_{i=0,1,\dots,N-1}$.
The initial state is chosen as 
\be
 |+\rangle=\frac{1}{\sqrt{N}}\sum_{i=0}^{N-1}|i\rangle,
\ee
and the Hamiltonian changes from $\hat{H}(0)=1-|+\rangle\langle+|$
to $\hat{H}(t_{\rm f})=1-|0\rangle\langle0|$.
In the adiabatic quantum computation, the Hamiltonian is changed slowly
and the system goes from the initial state to the target state $|0\rangle$.

The counterdiabatic Hamiltonian is written as 
\be
 \hat{H}_{\rm cd}(t)=\frac{i}{2}\sqrt{\frac{N}{N-1}}\dot{\theta}(t)\left(
 |0\rangle\langle+|-|+\rangle\langle0|\right), 
\ee
 where 
\be
 \theta(t)=\arctan\frac{\frac{\sqrt{N-1}}{N}A(t)}{\frac{1}{2}\left[
 \left(1-\frac{2}{N}\right)A(t)-B(t)\right]}.
\ee
For $\Theta(t)=\arccos|\langle 0|\psi(t)\rangle|$, 
the bounds are calculated as 
\be
 && \Theta_{\rm u}(t)=\frac{\pi-\theta(0)}{2}, \\
 && \Theta_{\rm l}(t)=\frac{\pi-\theta(0)}{2}-(\theta(t)-\theta(0)).
\ee
The upper bound is time independent, which shows that 
$\Theta$ does not exceed the initial value.
The lower bound shows that we can find the minimum time
for the time-evolved state to reach the target state.
The minimum time $t_{\rm min}$ is obtained 
by solving $\theta(t_{\rm min})=(\pi+\theta(0))/2$.

\begin{figure}[t]
\begin{center}
\includegraphics[width=0.80\columnwidth]{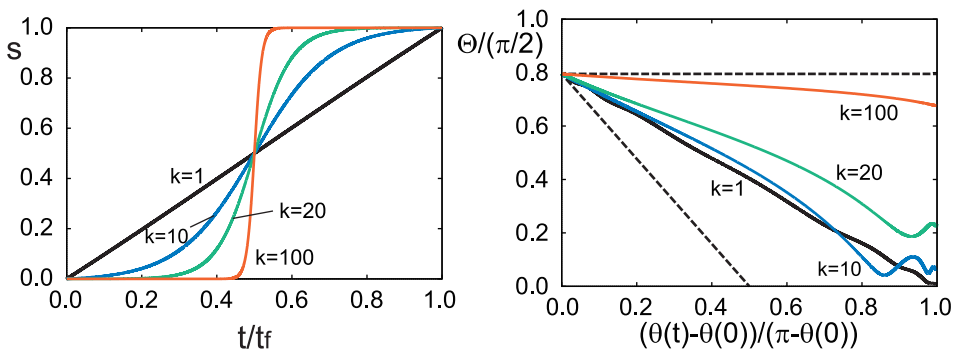}
\caption{Quantum speed limit of the Grover Hamiltonian 
with $N=10$, $A(0)t_{\rm f}=20$, and the protocol 1 in the left panel. 
The right panel represents 
$\Theta$ (solid lines) and $\Theta_{\rm u,l}$ (dashed lines).
}
\label{fig03}
\end{center}
\end{figure}

\begin{figure}[t]
\begin{center}
\includegraphics[width=0.80\columnwidth]{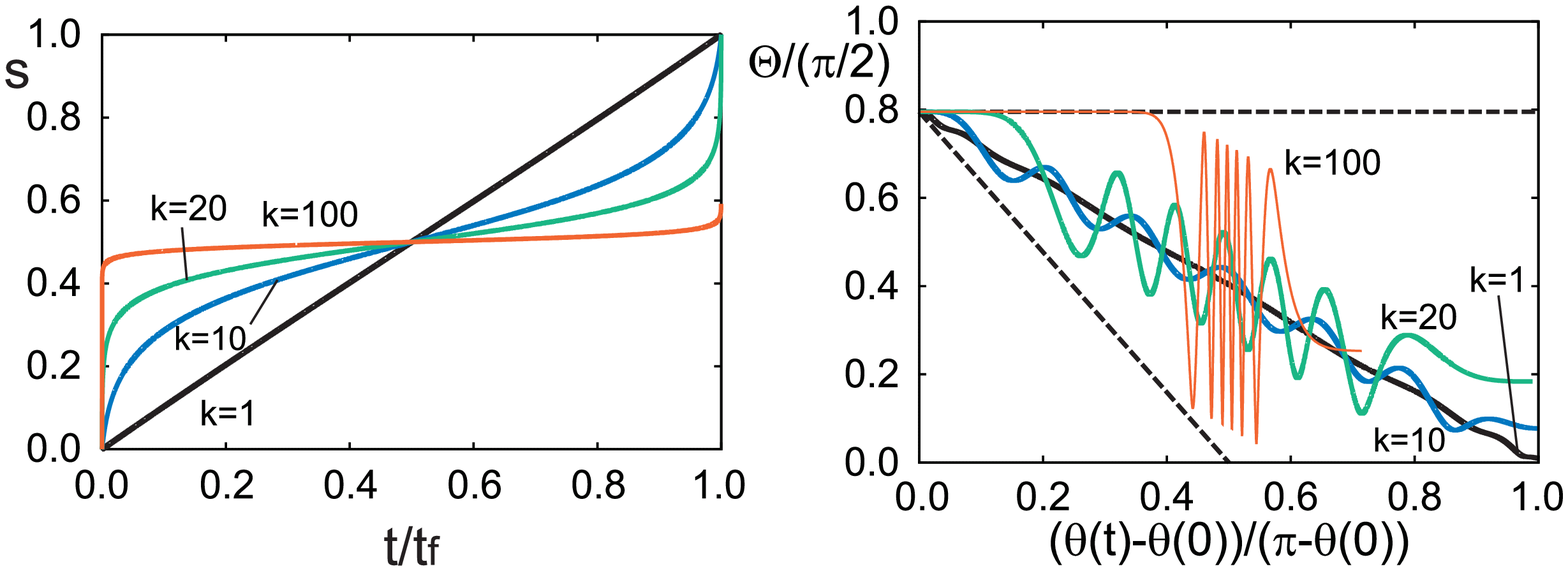}
\caption{Quantum speed limit of the Grover Hamiltonian 
with $N=10$, $A(0)t_{\rm f}=20$, and the protocol 2 in the left panel. 
The right panel represents 
$\Theta$ (solid lines) and $\Theta_{\rm u,l}$ (dashed lines).
}
\label{fig04}
\end{center}
\end{figure}

As a typical choice of the protocol in
the adiabatic quantum computation, we use 
$A(t)=A(0)(1-s(t/t_{\rm f}))$ and $B(t)=A(0)s(t/t_{\rm f})$~\cite{Takahashi19}.
The function $s(\tau)$ is an increasing function satisfying 
$s(0)=0$ and $s(1)=1$.
We choose two types of the protocol as 
\be
 s(\tau)=\left\{\ba{ll}
 \displaystyle 
 \frac{1-\rme^{-k\tau}}{(1-\rme^{-k/2})(1+\rme^{-k(\tau-1/2)})} & 
 \mbox{protocol 1}\\
 \displaystyle 
 \frac{1}{k}\ln\frac{(\rme^{k/2}-1)\tau+1}{1-(1-\rme^{-k/2})\tau} & 
 \mbox{protocol 2}
 \ea\right.,
\ee
where $k\ge 1$.
In both cases, $s(\tau)\sim\tau$ for $k=1$.
This linear protocol is frequently used in the adiabatic quantum computation.
As we increase $k$, $s(\tau)$ changes rapidly 
and the adiabaticity condition is broken.
We plot the protocol 1 in the left panel of figure \ref{fig03} and 
the protocol 2 in the left panel of figure \ref{fig04}.

The results are shown in the right panel 
of figure \ref{fig03} for the protocol 1 
and of figure \ref{fig04} for the protocol 2. 
In the present choice of parameters $N=10$ and $A(0)t_{\rm f}=20$, 
$\Theta$ is well described by the adiabatic approximation 
for small $k$ and shows a nonadiabatic behavior for large $k$.
When $k$ is a large value for the protocol 2, the result is rapidly 
oscillating and the amplitude of the oscillation is tightly bounded 
by $\Theta_{\rm u,l}$.
This result shows that $\Theta_{\rm u,l}$ 
are not overestimation and give reasonable bounds. 
As in the result of the previous section, 
the quantum speed limit is useful when we consider nonadiabatic regime.


\section{Periodically-oscillating system: bound optimization}
\label{floquet}

The third example treats a periodically-oscillating Hamiltonian $\hat{H}(t)$.
Generally speaking, the standard quantum speed limit bound becomes loose 
when the Hamiltonian changes back and forth.
We discuss how this problem is improved by using a proper reference state.

As a reference Hamiltonian, we use 
a time-independent Hamiltonian $\hat{H}_{\rm ref}$.
This choice allows us to evaluate the bound easily.
When $\hat{H}(t)$ changes rapidly, 
we can obtain the effective time-independent Hamiltonian 
from the Floquet--Magnus expansion~\cite{Magnus, BCOR}.
The effective Hamiltonian can be utilized as a reference one.

To obtain a tractable result, we use the exactly-solvable Hamiltonian
\be
 \hat{H}(t)=\frac{1}{2}\left(\ba{cc} \Delta & h\rme^{-i\omega t} \\
 h\rme^{i\omega t} & -\Delta \ea\right), \label{Hpo}
\ee
where $\Delta$, $h$, $\omega$ are positive parameters.
When $\omega$ is small, the adiabatic approximation gives a reasonable result
and the opposite limit, large $\omega$, is described by the Floquet--Magnus
expansion.
Here, we consider the large $\omega$ case.
The Floquet--Magnus expansion gives a time-independent Hamiltonian 
\be
 \hat{H}_{\rm ref}=\frac{1}{2}
 \left(\ba{cc} \Delta_{\rm ref} & h_{\rm ref} \\
 h_{\rm ref} & -\Delta_{\rm ref} \ea\right). \label{Href}
\ee
Up to the second order in $1/\omega$, we obtain 
\be
 && \Delta_{\rm ref}= \Delta-\frac{h^2}{2\omega}-\frac{\Delta h^2}{\omega^2}, \\
 && h_{\rm ref}= -\frac{\Delta h}{\omega}
 -\frac{\Delta^2 h}{\omega^2}+\frac{h^3}{2\omega^2}.
\ee
The reference state is obtained as 
$|\psi_{\rm ref}(t)\rangle=\rme^{-i\hat{H}_{\rm ref}t}|\psi(0)\rangle$.

\begin{figure}[t]
\begin{center}
\includegraphics[width=0.80\columnwidth]{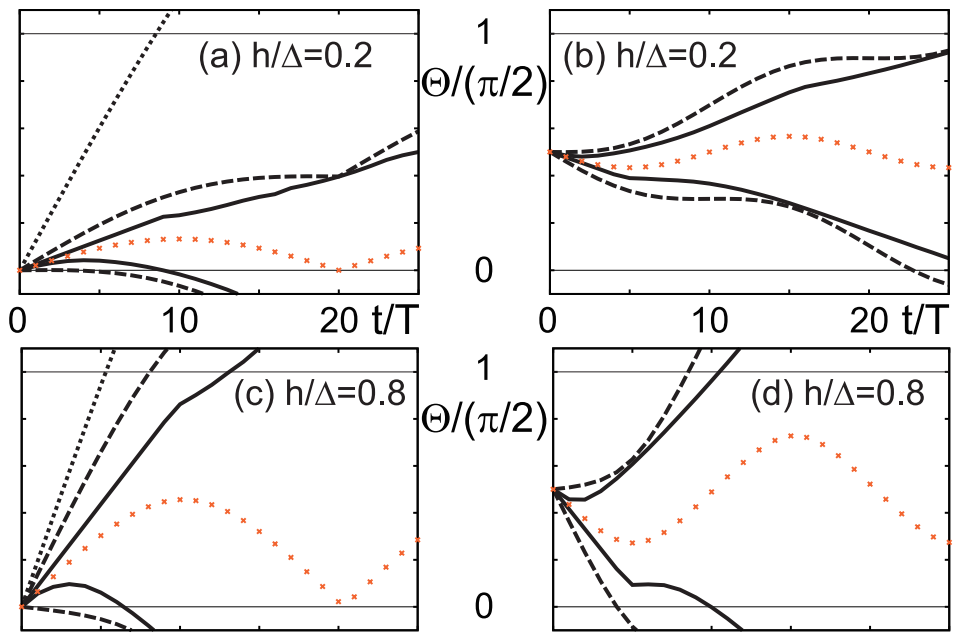}
\caption{Quantum speed limit of the periodically-oscillating Hamiltonian 
in equation (\ref{Hpo}).
$\Theta$ (points) is plotted as a function of $t$ at each panel.
$T=2\pi/\omega$ represents the period of the oscillation and 
we take $\omega/\Delta=20$.
The parameter $h/\Delta$ is $0.2$ for the panel (a) and (b) 
and $0.8$ for (c) and (d).
We use $|\psi_{\rm target}\rangle=|\psi(0)\rangle$  
for the panel (a) and (c), 
and equation (\ref{y}) for (b) and (d).
The bounds are estimated by three ways.
The standard quantum speed limit is represented by the dotted line.
The dashed lines represent $\Theta_{\rm u,l}$ 
in the case that the Floquet state is chosen for the reference state.
The solid lines represent $\Theta_{\rm u,l}$ 
for the optimized reference state.
}
\label{fig05}
\end{center}
\end{figure}

We start the time evolution from 
the eigenstate of the Hamiltonian $\hat{H}(0)$
with the positive eigenvalue.
When we choose the initial state 
as the target state $|\psi_{\rm target}\rangle=|\psi(0)\rangle$,
$\Theta(t)=\arccos |\langle\psi(0)|\psi(t)\rangle|$ changes from 
the initial value $\Theta(0)=0$ 
as we show crossed points in the panels (a) and (c) of figure \ref{fig05}.
In this case, the lower bound $\Theta_{\rm l}(t)$ (lower dashed line) 
is calculated to give 
negative values at any $t$ and we cannot obtain a useful result.
On the other hand, 
the upper bound $\Theta_{\rm l}(t)$ (upper dashed line) 
gives a reasonable result which is smaller than the bound obtained 
in the standard quantum speed limit (dotted line).

The negative lower bound is due to the choice of the target state.
In the panels (b) and (d) of figure \ref{fig05}, 
we show the result in the case of the target state 
\be
 |\psi_{\rm target}\rangle = 
 \frac{1}{\sqrt{2}}\left(\ba{c} 1 \\ i \ea\right). \label{y}
\ee 
In this case, the lower and upper bounds are between $0$ and $\pi/2$ 
at not too large $t$.

The use of the Floquet--Magnus expansion does not necessarily give 
a tight bound.
Although the reference state $|\psi_{\rm ref}(t)\rangle$
gives a good approximation to the time-evolved state $|\psi(t)\rangle$,
the variance $\sigma$ is not necessarily optimized.
We choose the reference Hamiltonian in equation (\ref{Href}) and
the parameters $\Delta_{\rm ref}$ and $h_{\rm ref}$ are optimized
so that $\Theta_{\rm u}$ is minimized and $\Theta_{\rm l}$ is maximized
at each $t$.
The  result is shown by the solid lines in figure \ref{fig05}.
The bounds are useful at transient times, several periods of the oscillation.
It is remarkable to notice that, in the panels (a) and (c), 
the lower bound gives a positive value at a small $t$.

\section{Summary and perspectives}
\label{summary}

We have discussed lower and upper quantum speed limits.
The inequalities are represented by using three quantum states.
The choices of the reference state and the target state are arbitrary
and we can considerably improve the standard bound.

We stress two important points in deriving the improved inequalities:
the triangle inequality in equation (\ref{tri}) and 
the estimation of the bound by the method in \cite{ST, HT}.
The triangle inequality is represented by using 
the density operator and the general distance measure 
and is generally satisfied without any approximation.
Although we have treated pure quantum states in this paper,
our method is applicable to broader classes of quantum and classical systems
in cases that the distance measure is defined.
In open quantum and classical systems, speed limits are derived 
in some situations, and 
it would be a straightforward task to apply the method developed in
the present paper.

It is also an interesting problem to apply the present method to 
systems with many degrees of freedom.
The system energy is typically proportional to the system size $N$,
which implies that the variance $\sigma$ is proportional to $N$.
The overlap between two different quantum states 
$|\psi_1\rangle$ and $|\psi_2\rangle$ has a form 
$|\langle\psi_1|\psi_2\rangle|\sim \rme^{-Ng}$ and 
the quantum speed limit inequality becomes a trivial relation.
Although it is possible to find a bound for 
the rate function $g=-\frac{1}{N}\ln|\langle\psi_1|\psi_2\rangle|$ \cite{ST},
$g$ is not a distance measure and we cannot use the triangle inequality.
In spite of this problem, 
we still have a possibility to find a meaningful result 
by using arbitrariness on the choice of the reference state.
It will be an interesting future problem.

\section*{Acknowledgments}

The author is grateful to Ken Funo and Takuya Hatomura for 
useful discussions and comments. 
This work was supported by JSPS KAKENHI Grants 
No. JP20K03781 and No. JP20H01827.

\section*{References}


\end{document}